\documentstyle[twoside,epsf,fleqn,espcrc2]{article}

\def\bc{\begin{center}}
\def\ec{\end{center}}
\def\be{\begin{equation}}
\def\ee{\end{equation}}
\newcommand{\beq}{\begin{equation}}
\newcommand{\eeq}{\end{equation}}
\newcommand{\beqn}{\begin{eqnarray}}
\newcommand{\eeqn}{\end{eqnarray}}

\def\vdir{v\kern-7.8pt\Big{/}}
\def\pdir{p\kern-7.8pt\Big{/}}

\title{New Results From Lattice QCD: Non--Perturbative Renormalization and
Quark Masses.}

\author{V.~Gim\'enez\address{Dep. de Fisica Teorica and IFIC, 
Univ. de Valencia,\\
Dr. Moliner 50, E-46100, Burjassot, Valencia, Spain.}%
\thanks{Invited talk at the QCD98 Conference. Work done in collaboration 
with L.~Giusti, F.~Rapuano and M.~Talevi.}
}

\begin{document}

\begin{abstract}
For the first time, we compute non--perturbatively, i.e.~without lattice perturbation theory,
the renormalization constants of two--fermion operators in the quenched 
approximation at $\beta=6.0$, $6.2$
and $6.4$ using the Wilson and the tree--level improved SW--Clover actions. We apply these renormalization constants to fully non--perturbatively 
estimate quark masses in the $\overline{MS}$ scheme from lattice simulations of both the 
hadron spectrum and the Axial Ward Identity in the quenched approximation.
Some very preliminary unquenched Wilson results obtained from the gluon
configurations generated by the T$\chi$L Collaboration at $\beta=5.6$ and $N_{f}=2$ 
are also discussed.
\end{abstract}

\maketitle

\section{Introduction.}
\label{introduction}

The values of the quark masses are very important for phenomenology.
Quark masses are free parameters of the QCD Lagrangian which cannot be
determined within QCD from theoretical considerations only. Since quarks are
confined inside hadrons, they are not observed as physical particles. Therefore,
{\bf quark masses cannot be measured directly and their values depend on how they 
are defined from observations, i.e. they are scheme and scale
dependent quantities}. 
%Nevertheless, quark masses are very important
%parameters for phenomenology: for instance, $m_{s}$, the mass of the $s$ quark,
%enters the $\Delta I=1/2$ $K\rightarrow \pi\pi$ amplitude and the ratio
%$\epsilon'/\epsilon$; light quark masses are necessary to estimate the chiral
%condensate and $m_{u}$  is crucial to understand the strong $CP$ problem.

The QCD Lagragian has a chiral symmetry for vanishing quark masses that is 
spontaneously broken by dynamics and explicitly broken by quark masses. 
%When
%the $SU(3)_{L}\times SU(3)_{R}$ chiral symmetry for the light quarks $u$, $d$ and
%$s$ is spontaneosly broken, eight massless Goldstone bosons are produced: $\pi$,
%$K$ and $\eta$. These bosons acquire masses due to the explicitly breaking of 
%the chiral symmetry by quark masses. 
%The masses of the Goldstone bosons
%can be computed in a systematic expansion in the quark masses, known as 
%chiral perturbation theory ($\chi PT$), plus certain
%unknown non--perturbative parameters so that to lowest order one has:
The masses of the Goldstone bosons $\pi$,
$K$ and $\eta$ can be computed in a systematic expansion in the quark masses, 
the so--called chiral perturbation theory ($\chi PT$), plus certain
unknown non--perturbative parameters, $\mbox{\rm A, B, C}$, so that to lowest 
order one has:
\beqn
\label{chipt}
m_{PS}^{2} &=& \mbox{\rm C}\, (m_{1} + m_{2})\nonumber\\
m_{V} &=& \mbox{\rm A}\, +\, \mbox{\rm B}\, (m_{1} + m_{2})
\eeqn
where $m_{PS}$ ($m_{V}$) is the mass of a pseudoscalar (vector) meson in terms
of the quark masses $m_{1,2}$. As a consequence, $\chi PT$ can only estimate
ratios of quarks masses but {\bf not their absolute values}.
%ratios of quarks masses, which are scale independent,
%but {\bf not their absolute values} \cite{gasleut}
%The coefficients $\mbox{\rm A, B, C}$ are unknown
%and cannot be calculated using $\chi PT$, thus $\chi PT$ can only estimate
%ratios of quarks masses, which are scale independent,
%but {\bf not their absolute values} \cite{gasleut},
%\beq
%\frac{m_{u}}{m_{d}}\, =\,
%0.553(43)\;\;\;\;\;\frac{2\,m_{s}}{m_{u}+m_{d}}\, =\, 24.41(1.5)
%\eeq
Therefore, to determine the absolute normalization of quark masses, we have 
to use methods which go beyond $\chi PT$, such as QCD Sum Rules or Lattice
QCD. In this work, we will use Lattice QCD to compute the meson masses and 
then, by utilizing the $\chi PT$ results, extract the quark masses. As an
alternative procedure, we will also simulate the Axial Ward Identity on
the lattice to calculate quark masses. 
%Both methods and their results 
%are explained in detail below. 
We do not compute the bottom quark mass
which can be extracted from the Lattice HQET \cite{bmass}.

\section{The lattice definition of quark masses.}
\label{deflatt}

%Lattice QCD is QCD on a finite space--time grid of lattice spacing $a$.
%Since $a$ is the only parameter which carries the dimension of length,
%quark masses are calculated in units of $a$, $m\, a$. Hence, it is
%necessary to determine $a$ in order to convert the lattice results to
%physical units. Notice that the space--time discretization ensures,
%on the one hand, the regularization of ultraviolet infinities and thus
%it can be used as a regularization scheme with which one can perform
%perturbative calculations. On the other hand, correlation functions of 
%hadronic operators can be calculated by employing Monte Carlo methods
%and numerical simulations.

In this work, we will consider the Wilson and the tree--level improved 
SW--Clover lattice actions in Euclidean space. %\cite{wilsw}
They are
parameterized in terms of $(\beta,\kappa)$ which are the only tunable input
parameters related to the bare lattice coupling constant, $g_{0}$, and the bare
lattice quark mass, $m_{0}$, via $\beta=6/g_{0}^{2}$ and $\kappa=1/(2\, m_{0}
a\, +\, 8)$ respectively, with $a$ the lattice spacing.
As is well known, these actions contain a term proportional to $a$, the
so--called Wilson term, which is needed to avoid the fermion doubling
problem, but explictly breaks the $SU(N_{f})_{L}\times SU(N_{f})_{R}$ chiral
symmetry. On the lattice, the chiral properties of QCD are lost so that 
the chiral limit does not correspond to vanishing bare lattice quark masses 
and operators of different chirality can mix among themselves.
In spite of this, it is possible to find some lattice currents which are 
partially conserved and obey the Current Algebra in the limit $a\rightarrow 0$ \cite{boch}.\\
{\bf Vector Ward Identities (VWI)} are obtained by applying infinitesimal vector
flavour transformations to the quark fields of the lattice actions and 
exploting the invariance of the path integral under local changes of the
fermionic variables. Using them, it is easy to show that there is a lattice
vector current, $\tilde{V_{\mu}}$, which is both conserved in the
limit of degenerate quarks, i.e.~$Z_{\tilde{V}}=1$, and improved, 
i.e.~its matrix elements have errors of order $O(g_{0}^{2} a)$ only. For the
Wilson action, the lattice
VWI between on--shell hadronic states $\alpha$ and $\beta$ takes the form
\beq
\langle \alpha | \partial_{\mu} \tilde{V}_{\mu} | \beta  \rangle\, =\,
\frac{1}{2}\, \left(\frac{1}{\kappa_{2}} - \frac{1}{\kappa_{1}} \right)\, 
\langle \alpha | S | \beta  \rangle
\eeq
where $\kappa_{1,2}$ are the hopping parameter of the quarks of the hadron,
$S=\bar{q}_{2} q_{1}$ is the bare scalar density and $\partial_{\mu}$ is the
forward lattice derivative. By comparing with the continuum QCD VWI, one obtains
the relation between the renormalized lattice quark mass and the hopping parameter
\beq
m_{R} a\, =\, Z_{S}^{-1}\, m a\, \equiv\, Z_{S}^{-1}\, \frac{1}{2} \left(\frac{1}{\kappa} - \frac{1}{\kappa_{c}} \right) 
\eeq
where $\kappa_{c}$ is the critical value of the hopping parameter
corresponding to a vanishing quark mass, and $Z_{S}$ is the renormalization
constant of the scalar density. For the SW--Clover action, a similar relation can be
derived (see \cite{noi}). In order to determine $\kappa$, one could compute 
the ratio of the vector and scalar densities on the lattice but, unfortunately, 
the scalar density matrix elements turn out to be extremely noisy, preventing
any reliable analysis. Alternatively, one can calculate $\kappa$ by fixing
the mass of an hadron containing a quark of a given flavour to its experimental
value using eq.(\ref{chipt}). This is the {\it Spectroscopy method}.\\
{\bf Axial Ward Identities (AWI)} are obtained analogously by applying
axial--vector transformations. In this case, however, there is no partially
conserved lattice axial current either for the Wilson or SW--Clover action.
Therefore, any lattice definition of the axial current get renormalized, even
in the limit of vanishing bare quark masses. For degenerate quarks, close to the
chiral limit ($\kappa = \kappa_{c}$) and neglecting terms of order $O(a)$, one
obtains the AWI \cite{boch}
\beq
\label{awi}
Z_{A}\, \langle \alpha | \partial_{\mu} A_{\mu} | \beta  \rangle\, =\,
\left(\frac{1}{\kappa} - \frac{1}{\kappa_{c}} \right)\, \frac{Z_{P}}{Z_{S}}\,
\langle \alpha | P | \beta  \rangle
\eeq
where $P=\bar{q}_{2} \gamma_{5} q_{1}$ is the bare pseudoscalar density, 
$A_{\mu}=\bar{q}_{2} \gamma_{\mu} \gamma_{5} q_{1}$ the bare axial current and
$Z_{P}$ and $Z_{A}$ are their renormalization constants.
From (\ref{awi}), one can define the renormalized quark mass as
\beq
m_{R} a\, =\, \frac{Z_{A}}{Z_{P}}\, \frac{\rho a}{2}\, \equiv\, 
\frac{Z_{A}}{Z_{P}}\, \frac{a}{2}\, \frac{\langle 0 | \partial_{0} A_{0} |
\pi(\vec{p}=0) \rangle}{\langle 0 | P |
\pi(\vec{p}=0) \rangle}
\eeq
where 
$|\pi(\vec{p}=0)\rangle$ is a pseudoscalar meson state containing a quark of 
hopping parameter $\kappa$.
For the SW--Clover improved case, the only difference is that one has to
consider the ratio of improved operators and take the symmetric derivative.
This is the {\it Axial Ward Identity method} which does not use $\chi$PT.

From the VWI and the AWI some interesting properties of the renormalization 
constants can be obtained:
\begin{enumerate}
\item $Z_{V}$, where $V_{\mu}=\bar{q}_{2} \gamma_{\mu} q_{1}$, 
% is the local vector
%current, 
and $Z_{A}$ are finite functions of $g^{2}_{0}$ with 
$Z_{V}$, $Z_{A}\neq 1$.
\item $Z_{P}$ and $Z_{S}$ are logarithmically divergent in $\mu a$ 
but $Z_{S}/Z_{P}$ is a finite function of $g^{2}_{0}$ only with 
$Z_{S}/Z_{P}\neq 1$ for the Wilson
and SW--Clover lattice actions.
%\item For a continuum regularization that breaks chirality and for the Wilson
%and SW--Clover lattice actions, $Z_{S}\neq Z_{P}$.
\end{enumerate}

\section{Evaluation of the renormalization constants.}
\label{renolatt}

Matching between lattice and continuum is necessary to convert 
the lattice results to the continuum renormalization scheme one has chosen
to analyze the experimental data. The matching factors $Z$'s are called lattice
renormalization constants. At scales $a^{-1}=2-4$ GeV of our simulations,
one expects small non--perturbative effects on the $Z$'s so that lattice (LPT)
and continuum (CPT) perturbation theory may be used to calculate them.
In almost all cases, however, the $Z$'s are known only to one--loop and,
further, is well known that LPT corrections are large due to {\bf tadpole} 
diagrams.
%, which are lattice artefacts because they do not appear in CPT.
It is claimed that the so--called {\it Boosted perturbation theory} (BPT) 
can resum tadpole diagrams by replacing the bare lattice coupling 
constant with a non--perturbatively renormalized one,
$\alpha_{P}=g_{0}^{2}/(4\pi) P$, where $P$ is the plaquette expectation value,
so that the observables calculated with this coupling are closer
to their non--perturbative counterparts.
%One attempt to solve this problem is based on the observation by Parisi,
%formalized and extended by Lepage and Mackenzie \cite{boosted}: resum the tadpole
%diagrams by replacing the bare lattice coupling constant
%$\alpha_{o}=g^{2}/4\pi$ with a non--perturbatively renormalized one
%$\alpha_{P}=\alpha_{o}/P$, where $P$ is the plaquette expectation value
%which can be obtained through lattice simulations
% and in its perturbative
%expansion contain large tadpole contributions
%. The claim is that the observables
%calculated with this {\it Boosted perturbation theory} (BPT) are closer
%to their non--perturbative counterparts. 
Unfortunately, data show that
%even after the implementation of the BPT,
even after the implementation of the BPT, $O(\alpha_{s}^{2})$ corrections 
can be large, of the order of $10$ -- $50 \%$, depending on the quantity at
hand. Alternatively, one can use the non--perturbative renormalization 
technique proposed by Martinelli {\it et al} \cite{np}. The idea is 
%very simple and
%powerful: one imposes renormalization conditions directly on quark and
to impose renormalization conditions directly on quark and
gluon Green functions in a fixed gauge
%, say the Landau gauge,
with given off--shell external
states with large virtualities $\mu^{2}$, mimicking the continuum renormalization
procedure but performing all computations non--perturbatively on the lattice.
This procedure defines the Regularization Independent scheme (RI), also called
MOM. For example, the renormalized scalar density in the RI scheme is  
$\hat{P}^{\mbox{\rm RI}}(\mu)\, =\, Z_{P}(\mu,a)\, P(a)$
where $Z_{P}$ is fixed by imposing the renormalization condition
\beq
\label{mom}
Z_{P}(\mu,a)\, \langle p | P(a) | p \rangle \mid_{p^{2}=-\mu^{2}}\, =\, 
\langle p | P(a) | p \rangle_{o}
\eeq
where $\langle p | P(a) | p \rangle_{o}$ is the tree level matrix element.
Since (\ref{mom}) can be evaluated non--perturbatively through lattice
simulations, $Z_{P}$ is extracted without LPT. Notice that 
$Z_{P}$ is affected by possible
$O(a)$ effects and depends on the external states and on the gauge. This
dependence, however, will cancel in the final results at a given order
when combined with the continuum $\overline{MS}$--RI matching calculation.
For this method, which defines the same renormalized operators in all
regularization schemes, to work the following window in $\mu$ must exist
\beq
\Lambda_{QCD}\, <<\, \mu\, <<\, a^{-1}
\eeq
where the first term is required to avoid large higher--order
corrections in CPT while the second one is necessary to avoid $O(a)$
effects.
The evolution of a bilinear operator renormalized in the $RI$ scheme
$\hat{O}^{RI}_{\Gamma}(\mu)$ ($\Gamma=V,A,P$ or $S$) can be 
expressed in the form 
\be
\hat{O}^{RI}_{\Gamma}(\mu)=
\frac{c^{RI}_{\Gamma}(\mu)}{c^{RI}_{\Gamma}(\mu_{0})}\,
\hat{O}^{RI}_{\Gamma}(\mu_{0})
\ee
where $c^{RI}_{\Gamma}$ are the solutions of the RGE which are known up to
N$^{2}$LO \cite{noi2}.
%The renormalization group equation can now be used to study the evolution of
%a bilinear operator renormalized in the $\overline{MS}$ scheme, 
%$\hat{O}^{\overline{MS}}_{\Gamma}(\mu)$ ($\Gamma=V,A,P or S$) up to N$^{2}$LO 
%because the corresponding anomalous dimensions are known up to three loops. 
%By calculating the matching between the 
%$\overline{MS}$ and the $RI$ schemes up to $O(\alpha^{s}_{s})$ for these
%operators, we can obtain the N$^{2}$LO evolution in the $RI$ scheme which can 
%be expressed in the form
%\be
%\hat{O}^{RI}_{\Gamma}(\mu)=
%\frac{c^{RI}_{\Gamma}(\mu)}{c^{RI}_{\Gamma}(\mu_{0})}\,
%\hat{O}^{RI}_{\Gamma}(\mu_{0})
%\ee
Therefore, we can define a renormalization group invariant quantity
\be
Z_{\Gamma}^{RGI}(a)\equiv\frac{Z^{RI}_{\Gamma}(\mu a)}{c_{\Gamma}^{RI}(\mu)}
\ee
which, up to higher order terms in CPT, should be independent of $\mu$, in the
region in which CPT is valid $\mu \ge 2$ GeV, of the renormalization scheme,
of the external states and gauge invariant. In fig.1 and 2, we show our
preliminary results for the $Z$'s for the unquenched Wilson action (see
ref.\cite{tedeschi}). We can clearly
see a window in $\mu$ where the $Z$'s are scale independent. The
corresponding $Z$'s for the Wilson and SW--Clover actions in the quenched
approximation are similar, but somewhat less stable, and can be found in
ref.\cite{noi2}. We have used them to perform the first measurement of quark 
masses without LPT.

\section{The continuum $\overline{MS}$ quark masses.}
\label{msmasses}

Since the $\overline{MS}$--RI matching factor depends on 
typical scales of order $\mu\approx a^{-1}\approx 2-4$ GeV, it can be calculated
in continuum perturbation theory only. We consider two cases:\\
{\bf Perturbative calculation of the $Z$'s}, i.e.~one uses both BPT and CPT at NLO
\[
m^{\overline{MS}}(\mu) = U^{\overline{MS}}_{m}(\mu,\pi/a)\, \times \,
\left\{\begin{array}{ll}
1/Z_{S}^{\overline{MS}}& m \\
Z_{A}^{\overline{MS}}/Z_{P}^{\overline{MS}} & \frac{\rho}{2}
\end{array}
\right.
\]
where $U^{\overline{MS}}_{m}$ is the evolutor operator for the quark mass
at NLO in the continuum. The renormalization constants $1/Z_{S}^{\overline{MS}}$ and  
$Z_{A}^{\overline{MS}}/Z_{S}^{\overline{MS}}$ are known to $O(\alpha_{s})$
only (see ref.\cite{noi2} for details).\\
{\bf Non--Perturbative calculation of the $Z$'s}, i.e.~one uses CPT but the $Z$'s
are computed non--perturbatively as explained in section 3. In this case, a N$^{2}LO$ matching
can be performed
\beqn
&&m^{\overline{MS}}(\mu) = U^{\overline{MS}}_{m}(\mu,\mu')\, 
\left[1 + \frac{\alpha_{s}(\mu')}{4 \pi} c^{(1)} \right.\nonumber\\
&+&\left(\frac{\alpha_{s}(\mu')}{4 \pi}\right)^{2} c^{(2)} \left.\right]
\times \,
\left\{\begin{array}{ll}
1/Z_{S}^{RI}(\mu' a)& m \\
Z_{A}^{RI}/Z_{P}^{RI}(\mu' a) & \frac{\rho}{2}
\end{array}
\right.\nonumber
\eeqn
where $c^{(1)}$ and $c^{(2)}$ are $\overline{MS}$--RI matching constants (see
ref.\cite{noi2} for details).
\begin{figure}[t]

\vspace{9pt}
\begin{center}\setlength{\unitlength}{1mm}
\begin{picture}(55,37)
\put(12,-10){\epsfbox{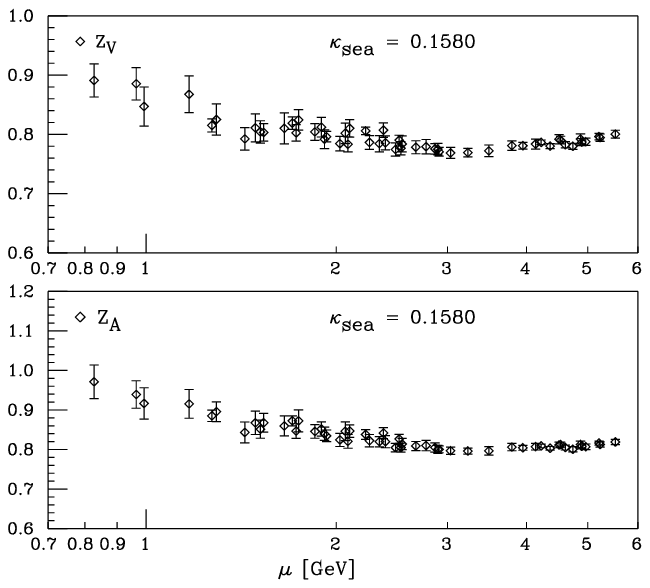}}
\end{picture}
\end{center}
\caption{\it{$Z_{V}$ and $Z_{A}$ for the unquenched Wilson action as a 
function of $\mu$ at $\kappa_{sea}=0.1580$.}}
\label{fig:fourfig1}
\end{figure}
\section{Results.}
\label{results}

\begin{figure}[t]

\vspace{9pt}
\begin{center}\setlength{\unitlength}{1mm}
\begin{picture}(55,40)
\put(12,-10){\epsfbox{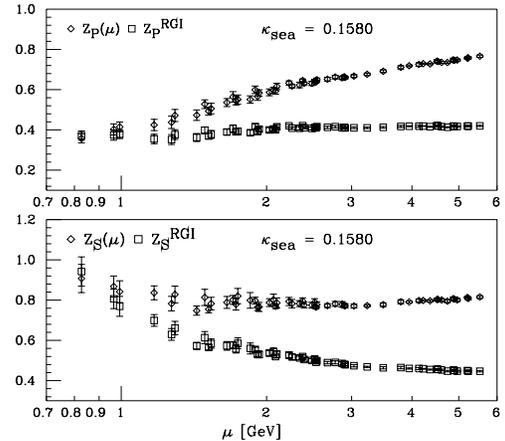}}
\end{picture}
\end{center}
\caption{\it{$Z_{P}$ and $Z_{S}$ for the unquenched Wilson action as a 
function of $\mu$ at $\kappa_{sea}=0.1580$.}}
\label{fig:fourfig2}
\end{figure}
 
We have analyzed about $1000$ quenched configurations for the Wilson action and
about $2000$ quenched configurations for the SW--Clover action in a variety of 
physical volumes (ranging from $16^{3}\times 32$ to $24^{3} \times 64$)
and lattice spacings corresponding to $\beta=6.0$, $6.2$ and
$6.4$. We refer the reader to the ref.\cite{noi} for technical details. For the
spectroscopy method, we find a resonable agreement between the quark masses
obtained with perturbative (PT) evaluated renormalization constants 
and non--perturbative (NP) ones and are compatible with previous determinations. 
This not the case, however, for the AWI method where the PT results are 
lower than the NP ones by more than two standard deviations. Only when 
we use non--perturbative renormalization constants, the spectroscopy and AWI
methods give consistent values of the quark masses, as it should be.
We think that the reason is that lattice perturbation theory fails to determine
$Z_{P}$ even if the Boosted recipes are implemented.
As for the charm quark mass, our results are less stables than for the light and
strange quarks. We think that  the reason is that there is a rather
large $O(m a)$ contamination in our data for the charm quark mass.
Finally, we want to stress that we cannot extrapolate to the continuum limit, 
even though we have simulated at three values of $\beta$, due to the small 
physical volume at $\beta=6.4$. Since our results at $\beta=6.0$ and $\beta=6.2$
are definitely stables within errors, i.e.~we do not see any dependence on $a$
in our results, we average the NP AWI results and obtain
our best estimates:
\beqn
\bar{m}^{\overline{MS}} (2\, \mbox{\rm GeV}) &=& 5.7 \pm 0.1 \pm 0.8\;\;  \mbox{\rm
MeV}\nonumber\\
m_{s}^{\overline{MS}} (2\, \mbox{\rm GeV}) &=& 130 \pm 2 \pm 18\;\;  \mbox{\rm
MeV}\nonumber\\
m_{c}^{\overline{MS}} (2\, \mbox{\rm GeV}) &=& 1662 \pm 30 \pm 230\;\;  \mbox{\rm
MeV}
\eeqn
where the first error is statistical and the second is an estimate of the
systematic error evaluated from the spread in the values of the quark masses
extracted from different mesons, lattices and chiral extrapolation methods.

\end{document}